\documentclass[twocolumn,aps,prl,groupedaddress]{revtex4}
\usepackage{epsfig,amssymb,amsmath}

\def\comment#1{}
\def\labell#1{\label{#1}}
\def\togli#1{}
\begin{document}
\title{Sub-Rayleigh Quantum Imaging} \author{Vittorio Giovannetti$^1$,
  Seth Lloyd$^{2}$, Lorenzo Maccone$^3$, and Jeffrey H. Shapiro$^2$}
\affiliation{$^1$NEST-CNR-INFM \& Scuola Normale Superiore, Piazza dei
  Cavalieri 7, I-56126, Pisa, Italy.\\$^{2}$MIT, Research Laboratory of
  Electronics, 77 Mass.~Ave., Cambridge, MA 02139, USA.\\
  $^3$QUIT, Dip.  Fisica ``A.  Volta'', Univ. Pavia, via Bassi 6,
  I-27100 Pavia, Italy.  }
\begin{abstract}
  No imaging apparatus can produce perfect images: spatial resolution
  is limited by the Rayleigh diffraction bound that is a consequence of the imager's
  finite spatial extent. We show some $N$-photon strategies that
  permit resolution of details that are smaller than this bound,
  attaining either a $1/\sqrt{N}$ enhancement
  (standard quantum limit) or a $1/{N}$ enhancement (Heisenberg
  limit) over standard techniques. In the incoherent
  imaging regime, the methods presented are loss resistant, because they
  can be implemented with classical-state light sources. Our results may be
  of importance in many applications: microscopy, telescopy,
  lithography, metrology, {\em etc.}
\end{abstract}
\maketitle 

Quantum effects have been used successfully to provide resolution
enhancement in imaging procedures. Among the many 
proposals that have been made~\cite{review}, arguably the most famous is the quantum
lithography procedure~\cite{litho}. All of these methods take advantage
of the fact that the de Broglie wavelength of a multi-photon light
state is much shorter than the photon's electromagnetic field
wavelength~\cite{chuangyama}: the light generation, propagation, and
detection can be performed at optical wavelengths, where it is simple
to manipulate, whereas the quantum correlations in the employed
states allow one to perform imaging at the much shorter de Broglie
wavelength. Such proposals are then based on light sources of
highly entangled or squeezed states, as entanglement or squeezing are
necessary to achieve efficient quantum enhancements~\cite{metrology}.
If, however, efficiency considerations are dropped, it is also
possible to employ classical-state light sources and post-selection at the
detection stage to filter desirable quantum states from the 
classical light~\cite{postsel}. In fact, in many practical situations
efficiency considerations do not play any role, as the quantum
enhancement is typically of the order of the square root of the number
of entangled systems~\cite{metrology}, whereas in practical situations
the complexity of generating the required quantum states has a much
worse scaling. Many post-selection imaging procedures employing only
classical light sources have been proposed and
analyzed~\cite{scullylith,korobkin,agarwal,boyd,zhang,wang,yablo,peer,shih,lugiato,jeff,jeffhorace,libroqim},
and cover a wide range of interesting situations.  Analogous methods
have been employed successfully also in fields not directly related to
imaging~\cite{altri}.

This Letter discusses how one can achieve a resolution enhancement
beyond what the apparatus' structural limits impose for conventional
imaging (i.e., the Rayleigh diffraction bound $x_R$). In particular we
show that employing appropriate light sources together with $N$-photon
coincidence photodetection at the output yields a resolution $\sim$$
x_R/\sqrt{N}$. A resolution $\sim$$x_R/N$ can also be obtained by
introducing, at the lens plane, a device that is opaque when it is illuminated by fewer 
than $N$ photons. The first type of enhancement---a
standard quantum limit for imaging---is an $N$-photon quantum
process, but it is roughly equivalent to the classical procedure of
averaging the arrival positions of $N$ photons that originate from  the same
point on the object.  The second type of enhancement---a Heisenberg
bound for imaging---is a quantum phenomenon that derives from
treating the $N$ photons as a single field of $N$-times higher frequency.
In the incoherent imaging regime, both methods presented
here can tolerate arbitrary amounts of loss at the expense of reduced efficiency but without sacrificing resolution.  

We start by reviewing some basics of conventional imaging. Then we
discuss coherent and incoherent sub-Rayleigh imaging procedures that
attain the standard-quantum limit, and finally we introduce our
approach to realizing the Heisenberg limit for imaging.

\paragraph{Rayleigh bound:--}Consider  monochromatic
imaging using a circular-pupil thin lens of radius $R$ and focal length $f$
that is placed at a distance $D_o$ from an object of surface area
$\cal{A}$, and at a distance $D_i$ from the image plane, where $1/D_0 + 1/D_i = 1/f$. 
In conventional
imaging, the object is illuminated by an appropriate (spatially coherent or incoherent) source
and the image plane distribution of the light intensity, corresponding to the probability of detecting a photon at each image-plane point $\vec{r}_i$, is recorded.  For photodetectors whose spatial-resolution area ${\cal S}$ and temporal-resolution time $\Delta t$ are sufficiently small, the preceding probability  satisfies $ P_1(\vec{r}_i)\simeq {(\eta {\cal S} c \Delta t)} \big\langle
E_i^{(-)}(\vec{r}_i,t) E_i^{(+)}(\vec{r}_i,t)\big\rangle$, where angular brackets denote
ensemble average over the illumination's state, $\eta$ is the
detector quantum efficiency, and $E^{(+)}=[E^{(-)}]^\dag$ is the
positive-frequency component of the electric field.  This field component obeys
$E^{(+)}_i(\vec{r}_i,t)=\int d^3\vec k\; {\cal
  E}_i^{(+)}(\vec{r}_i;\vec k)\;e^{-i {k} t/c}a(\vec k)$, where
$a(\vec k)$ is the field annihilation operator for the optical mode with
wave vector $\vec k$, and ${\cal E}_i^{(+)}$ is the solution to the
associated Helmholtz equation at the image plane. The latter can
be written in terms of the corresponding object-plane field ${\cal
  E}_o^{(+)}(\vec{r}_o,\vec k) = e^{i\vec k_t\cdot\vec{r}_o}$, where $\vec{k}_t$ is
the transverse component of $\vec{k}$, using classical imaging equations.  For monochromatic light in
the paraxial regime ${k}_t\ll{k}$, it follows that~\cite{goodman,born}
\begin{eqnarray} {\cal E}^{(+)}_i(\vec{r}_i;\vec k)=  \;
 \int \tfrac{d^2\vec{r}_o}{\cal A}
  \;A(\vec{r}_o) \; h(\vec{r}_i,\vec{r}_o)\;{\cal
    E}^{(+)}_o(\vec{r}_o;\vec k)
\labell{inputoutput}\;,
\end{eqnarray}
where $\vec{r}_i$ and $\vec{r}_o$ are two-dimensional vectors in the
image and object planes, $A(\vec{r}_o)$ is the object aperture
function~\cite{nota1}, and $h(\vec{r}_i,\vec{r}_o)$ is 
the point-spread function of the imaging apparatus given
by~\cite{shih,goodman,born}
\begin{eqnarray}
h(\vec{r}_i,\vec{r}_o)=
\tfrac{R^2 {k}^2 {\cal A}}{4 \pi D_oD_i}e^{i\vartheta}
{\rm somb}(R\; {k} \; |\vec{r}_o+\vec{r}_i/m|/D_o),
\labell{acca}\;
\end{eqnarray}
with ${\rm somb}(x)\equiv 2J_1(x)/x$ being the well known Airy
function, and $m = D_i /D_o$ being the image magnification factor.  In
Eq.~\eqref{acca}, $\vartheta$ is a phase factor which can be generally
neglected or compensated.

Incoherent imaging occurs when the object is illuminated by independent (monochromatic) 
beams propagating from all directions, whence
\begin{eqnarray}
  P_1^{(\text{inc}) } (\vec{r}_i)  = \tfrac{\eta {\cal S} c \Delta t} {2 \pi  {k}^2 {\cal A}}\; I_o
\int \tfrac{d^2\vec{r}_o}{\cal A}
  \;\Big|A(\vec{r}_o)
  \:h(\vec{r}_i,\vec{r}_o)\Big|^2
\labell{inco},
\end{eqnarray}
with $I_o\equiv \langle E_o^{(-)} E_o^{(+)}\rangle$ being the field
intensity on the object plane.  Coherent imaging prevails 
when collimated coherent-state illumination is employed, giving rise to 
\begin{eqnarray}
 P_1^{(\text{c}) } (\vec{r}_i) = {\eta {\cal S} c \Delta t} \;  I_o \;
\Big|\int \tfrac{d^2\vec{r}_o}{\cal A}
\;A(\vec{r}_o)\:
\:h(\vec{r}_i,\vec{r}_o)\Big|^2
\labell{conventional}\;.
\end{eqnarray}
When the lens radius $R$ is sufficiently large, Eqs.~(\ref{inco}) and (\ref{conventional})
produce inverted, magnified, perfect images of the object,
because $R^2{\rm somb}(R|\vec x|)\to 4\pi\delta^{(2)}(\vec x)$ for
${R\to\infty}$~\cite{goodman}.  For $R$ insufficient to reach this asymptotic behavior, the convolution
integrals in Eqs.~(\ref{inco}) and (\ref{conventional})
 produce blurred images. The amount of blurring can be gauged through the Rayleigh
diffraction bound: for a point source at $\vec{r}_o$ in the object plane, the resulting image-plane intensity is proportional to ${\rm somb}^2(R\,k\,|\vec{r}_o + \vec{r}_i/m|/D_o)$, which comprises a pattern of circular fringes in $\vec{r}_i$ that are centered on $-m\vec{r}_o$.  The radius of the first fringe~\cite{born},
\begin{eqnarray}
x_R\simeq 0.61\times 2\pi m D_o/(kR)\labell{rayldiff}\;,
\end{eqnarray}
about $-m\vec{r}_o$ encloses $\sim$84\% of the light falling on the image plane.  
Intuitively, the image of an extended object is then a weighted superposition of radius-$x_R$ circles of
centered about each $-m\vec{r}_o$.  This is the
Rayleigh diffraction bound; using conventional imaging techniques one cannot resolve details smaller than $x_R$.

\begin{figure}[h!]
{\footnotesize a}  \epsfxsize=.46\hsize\leavevmode\epsffile{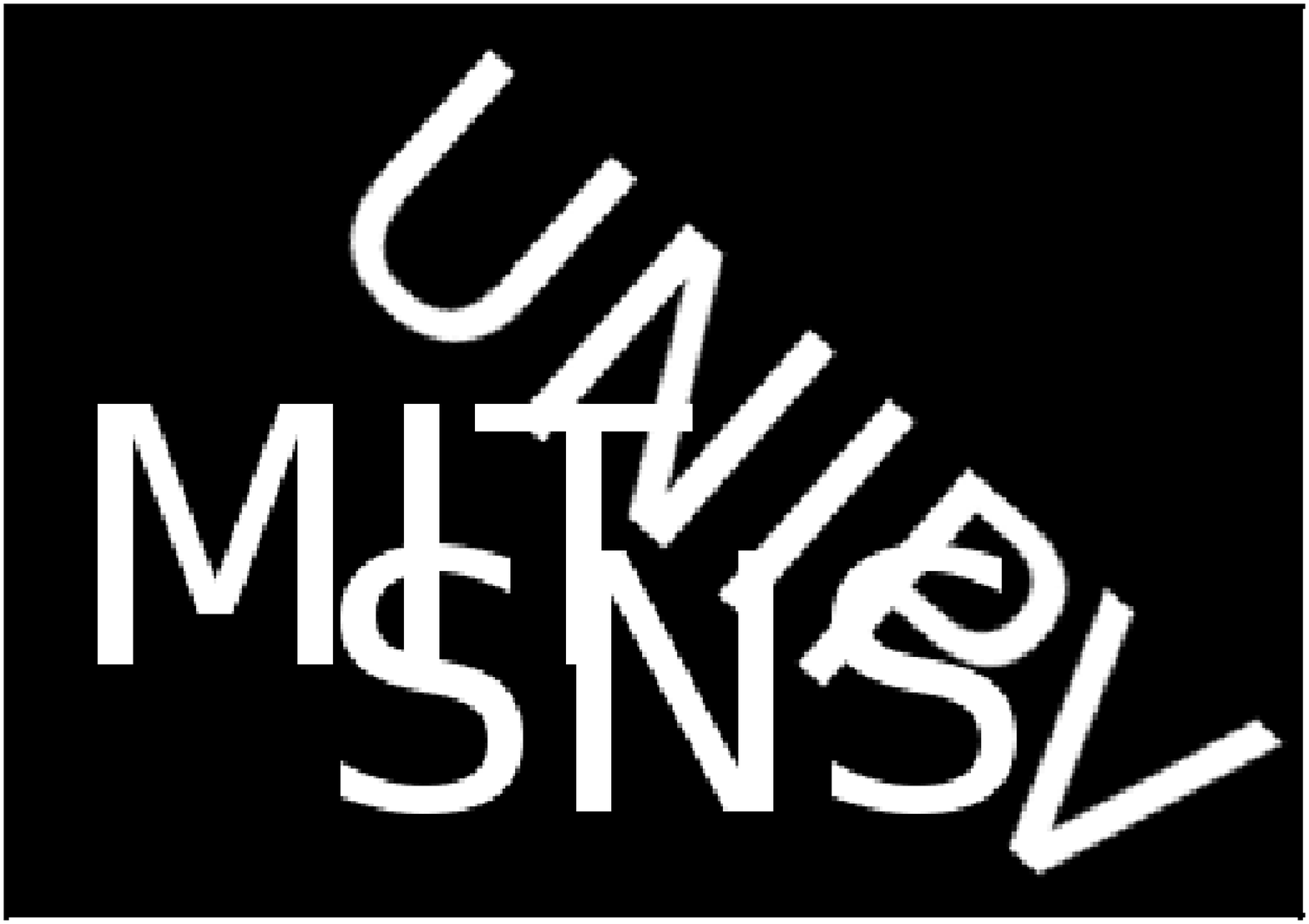}
{\footnotesize b}  \epsfxsize=.46\hsize\leavevmode\epsffile{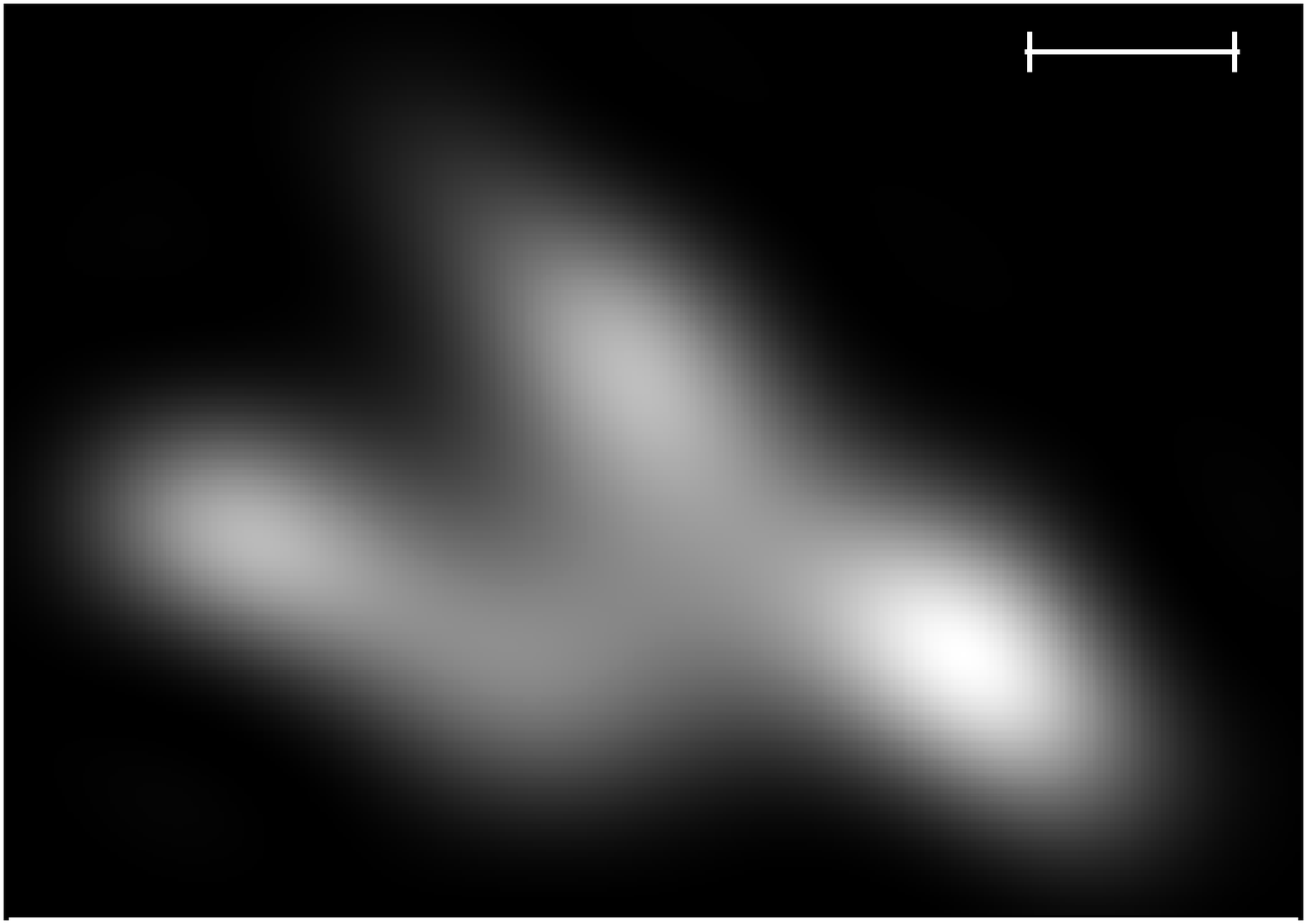}
\begin{center}
{\footnotesize c}  \epsfxsize=.46\hsize\leavevmode\epsffile{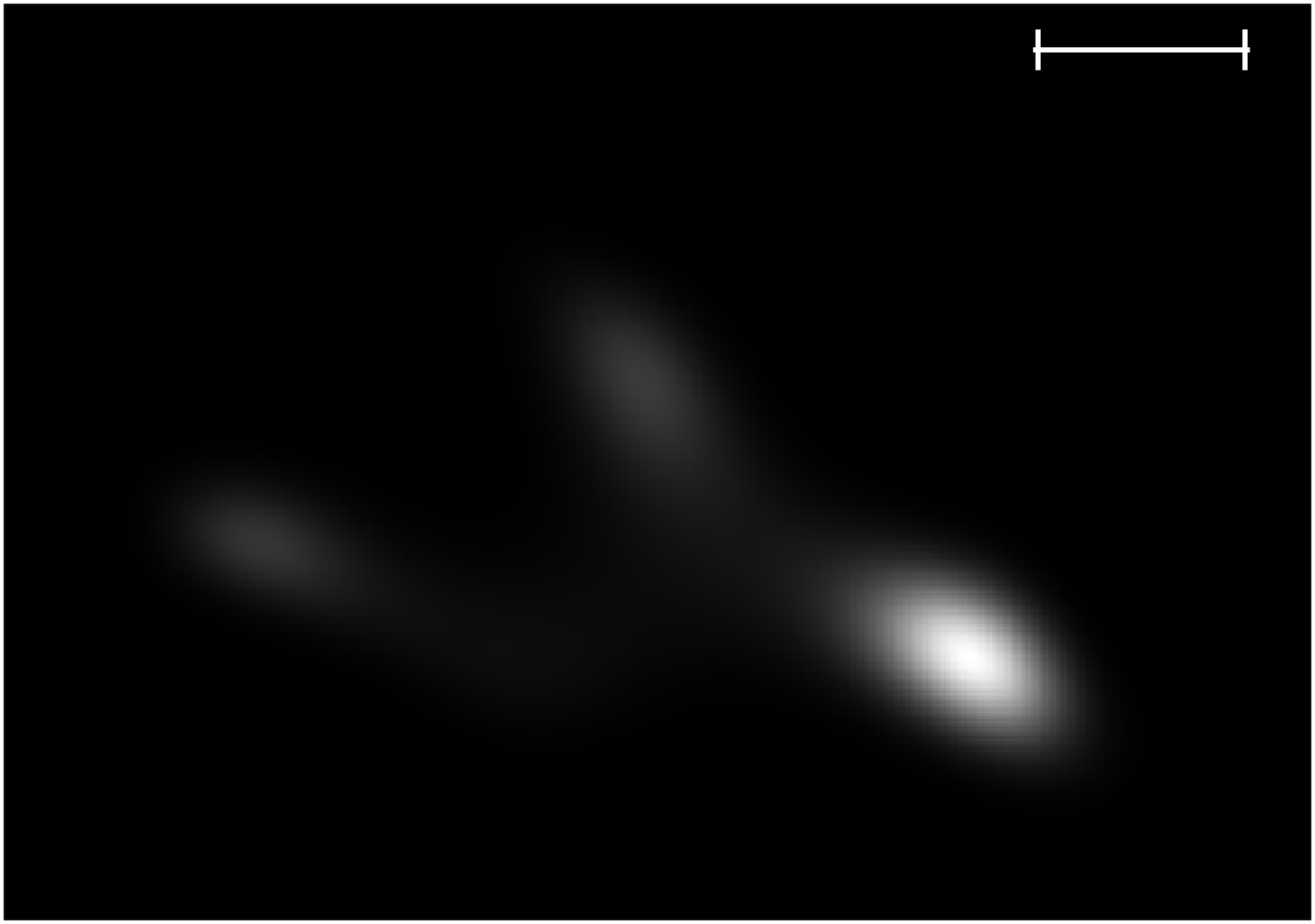}
{\footnotesize d}  \epsfxsize=.46\hsize\leavevmode\epsffile{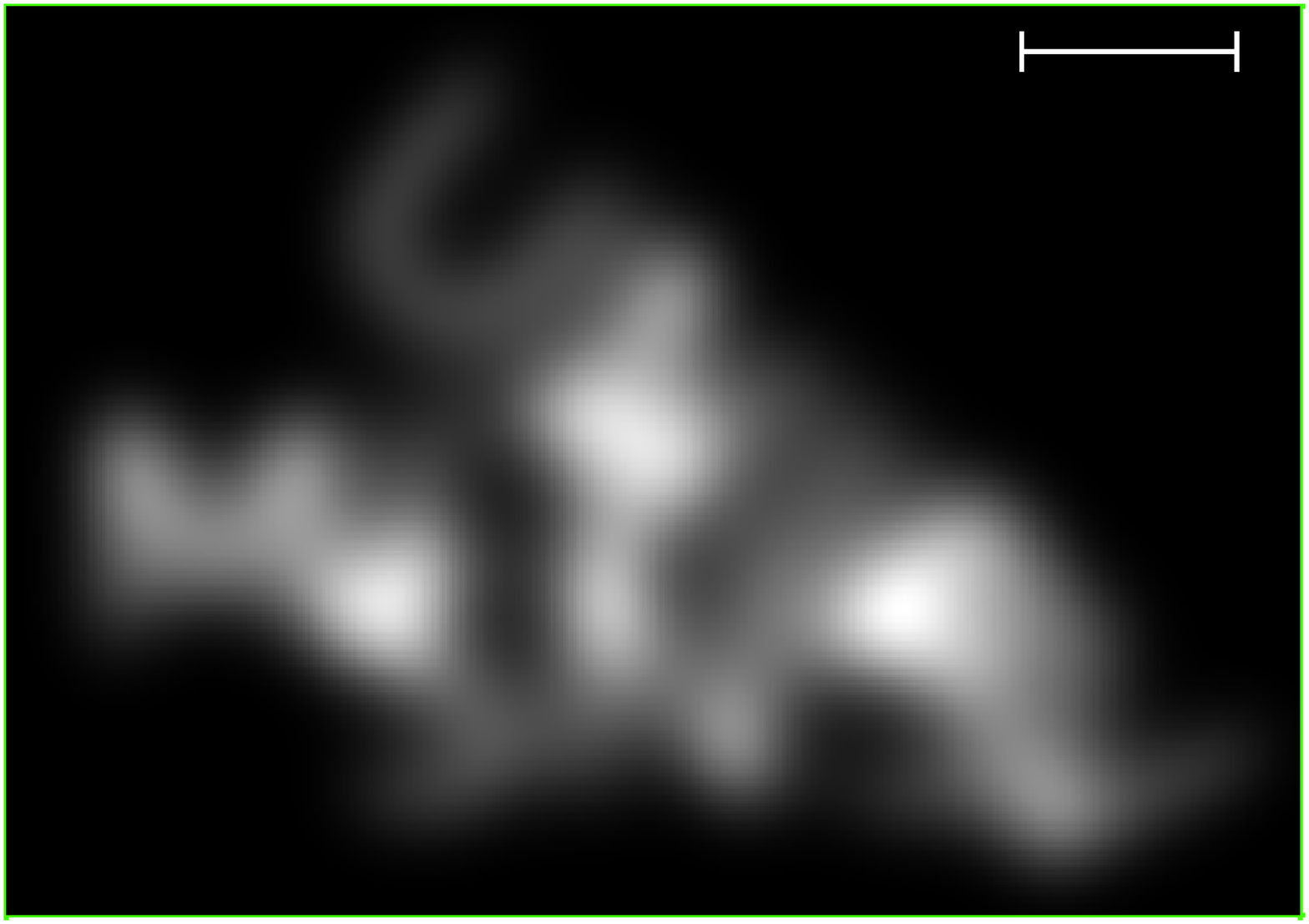}
\end{center}
\begin{center}
{\footnotesize e}  \epsfxsize=.46\hsize\leavevmode\epsffile{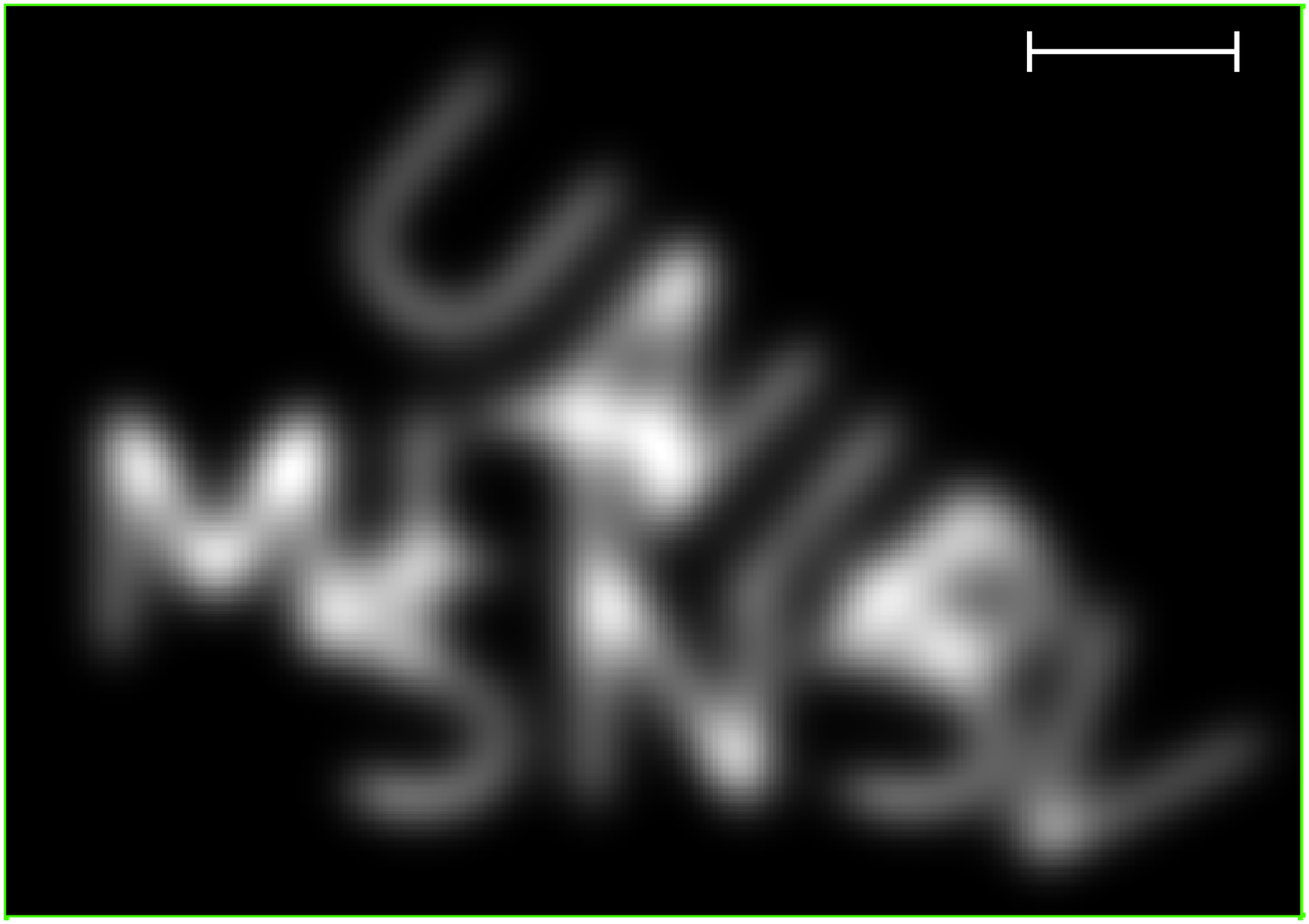}
{\footnotesize f}  \epsfxsize=.46\hsize\leavevmode\epsffile{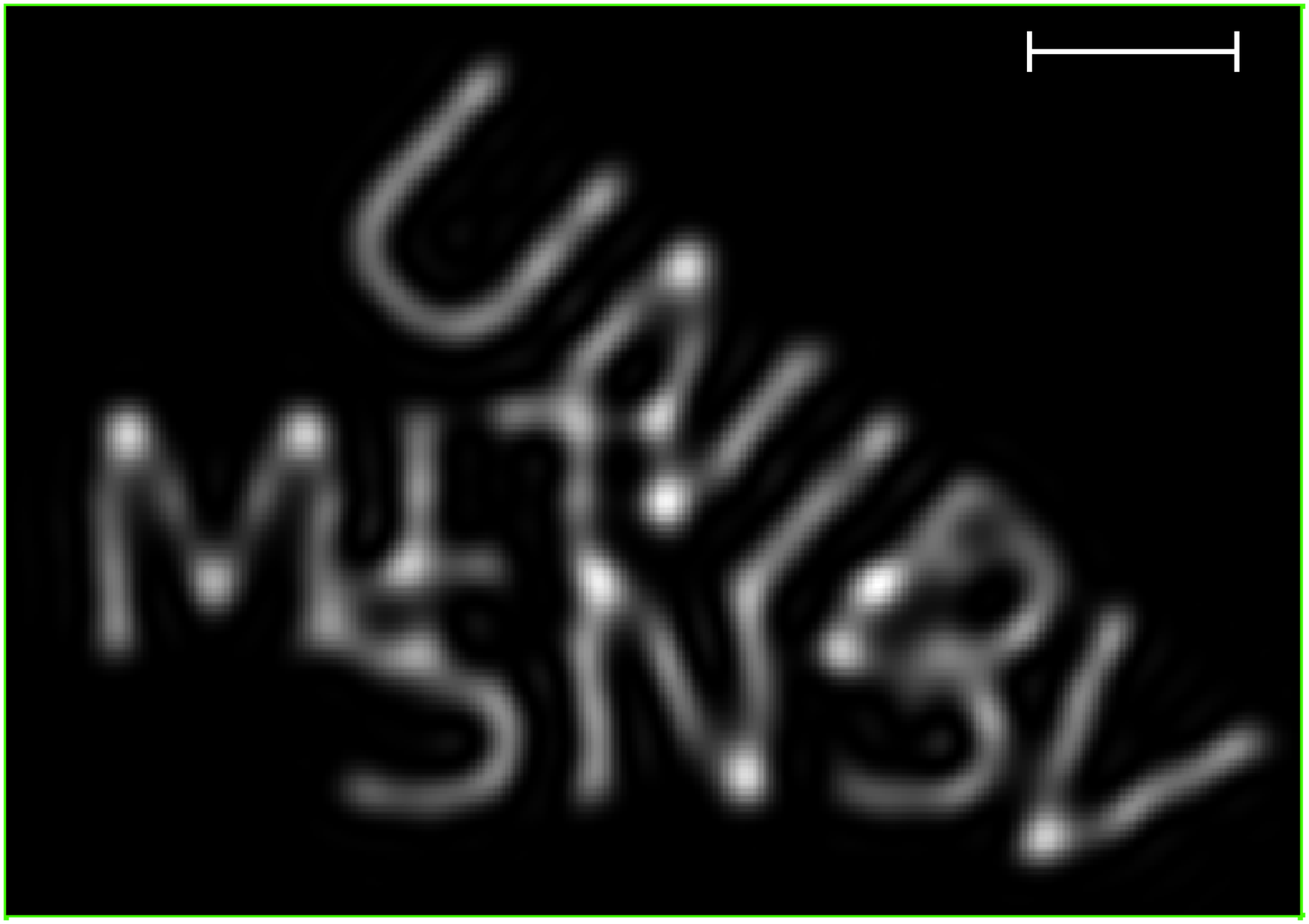}
\end{center}
\caption{Comparison of conventional, standard quantum limit, and Heisenberg limit coherent-imaging for a case in which the Rayleigh bound is unable to resolve any object details.  (a)~Object to be imaged.
  (b)~Conventional coherent image computed for $D_o/R=250$,
  $m=1$, and $k=6000$.  The length scale for $\vec k$ is
  given by the image width, and the segment in the top right corner is
  the Rayleigh bound $x_R$. Owing to diffraction, no object details are discernible in the image. (c)~$N$-fold
  coincidence detection ($N=5$) with coherent illumination,
  i.e., the reconstruction through Eq.~(\ref{nfold}). No resolution
  enhancement over the previous case occurs despite the $N$th-order compression of the point-spread function.
  (d) Standard-quantum-limit reconstruction with illumination by the superposition of
  Fock states from Eq.~(\ref{oak}) with $N=5$
 and $\Delta k_t=600$. Sub-Rayleigh resolution is present.  (e) The same as (d) except that $N = 10$; more resolution enhancement occurs.  (f)~Heisenberg-limited coherent reconstruction from Eq.~(\ref{enne}) with $N=5$; still further enhancement is evident.}
\labell{f:drago}
\end{figure}

\paragraph{Standard quantum limit:--}The main idea of sub-Rayleigh imaging is to use an appropriate light source and to replace intensity measurement with 
spatially-resolving $N$-fold coincidence detection strategies.
Specifically we will focus  on the probability of detecting $N$ photons at
position $\vec{r}_i$ on the image plane~\cite{mandel}, i.e.,
\begin{eqnarray}
  P_N(\vec {r}_i)\simeq \tfrac{( \eta {\cal S} c
   \Delta t)^N}{N!} 
\big\langle \big[E_i^{(-)}(\vec{r}_i,t)\big]^N
\big[E_i^{(+)}(\vec{r}_i,t)\big]^N\big\rangle
\labell{mandel}\;,
\end{eqnarray}
which can be accomplished by means of doppleron
absorbers~\cite{doppleron}, photon-number resolving detectors, or
$N$-fold  coincidence counting.  The last two approaches are more
convenient than the first, as they exploit the full photon statistics so that the 
$N$ value need not be predetermined.  Note that multi-photon detection
alone does not guarantee sub-Rayleigh performance. In fact, for
the coherent imaging of Eq.~(\ref{conventional}), $N$-photon detection
gives
\begin{eqnarray}
  P_N(\vec{r}_i)=\frac{1}{N!} \;
  {\left[P_1^{(\text{c})}(\vec{r}_i)\right]^N}\;.
  \labell{nfold}
\end{eqnarray}
Here, the factor of $N$ in the exponent gives an $N$-fold compression of
the fringes in the point-spread function.  This compression, however, 
is not intrinsically quantum.   It amounts to taking 
the $N$th power of the light intensity, which is simply a
classical post-processing of the signal in Eq.~(\ref{conventional}).  Thus no resolution enhancement is obtained through simple $N$-photon detection, see Fig.~\ref{f:drago}(c).

As our first example of a source that can be used to beat the Rayleigh
bound, consider an input state that is the superposition of
$N$-photon Fock states that have been focused to a small area $s\equiv (\pi
\Delta k_t^2)^{-1} \ll {\cal A}$ centered at positions $\vec{r}_o$ on
the object plane, viz., 
\begin{eqnarray}
 |\psi\rangle \equiv  \tfrac{1}{\sqrt{{\cal  M}}} 
 \int {d^2\vec{r}_o}\;|N\rangle_{\vec{r}_o} \;, \;\;
|N\rangle_{\vec{r}_o}\equiv\tfrac{1}{\sqrt{N!}}
\big[{b}^\dag(\vec{r}_o) \big]^N |0\rangle,
 \labell{oak}
\end{eqnarray}
where $b(\vec{{r}_o})$ is the annihilator of the associated localized
spatial mode~\cite{DEFB} and ${\cal
  M}\simeq \tfrac{16 \pi {\cal A}}{\Delta k_t^2}$ is a normalization
constant.  Inserting this state into Eq.~(\ref{mandel}), we find
\begin{eqnarray}
&&P_N(\vec{r}_i) \simeq \tfrac{\Delta k_t^2 {\cal A}}
 {16 \pi  } \; \xi^N \;
 \Big| \int \tfrac{d^2\vec{r}_o}{\cal A} \;
Q^N(\vec{r}_i,\vec{r}_o) \Big|^2
\labell{vv}\;, \\\nonumber
&&Q(\vec{r}_i,\vec{r}_o)\equiv \int \tfrac{d^2\vec{r}}{\cal A} \; \;A(\vec{r}) \;
{h}(\vec{r}_i,\vec{r})\;F_{\Delta k_t}(|\vec{r}_o-\vec{r}|)\;, 
\end{eqnarray}
where $F_{\Delta k_t}(x)\equiv \pi\Delta k_t^2 {\cal A} \;
{\rm somb}(\Delta k_t\;x/2)$, and  $\xi \equiv \eta \tfrac{\Delta
  \omega \Delta t}{\pi \Delta k_t^2 {\cal A}} \frac{{\cal S}}{\cal A}$
is a dimensionless quantity that is typically very small because of the
monochromatic ($\Delta \omega \Delta t \ll1$) and focusing assumptions
($\pi \Delta k_t^2 {\cal A} \gg 1$). Equation~(\ref{vv}) can be
simplified by assuming $D_o/R\gg {k}/\Delta k_t$, which implies that each number state in the superposition is focused to a spot much smaller than the object-plane Rayleigh limit of the lens.  In
this case ${h}$ can be extracted from the
integral yielding $Q(\vec{r}_i,\vec{r}_o)\simeq {h}
(\vec{r}_i,\vec{r}_o)\tilde A(\vec{r}_o)$ with $\tilde
A(\vec{r}_o)\equiv \int \tfrac{d^2\vec{r}_o}{\cal A}
A(\vec{r})F_{\Delta k_t}(|\vec{r}_o-\vec{r}|)$.  Now Eq.~(\ref{vv})
becomes
\begin{eqnarray}
P_N^{(\text{c})}(\vec{r}_i)\simeq \tfrac{\Delta k_t^2 {\cal A}}
 {16 \pi  } \; \xi^N  \Big|\int
\tfrac{d^2\vec{r}_o}{\cal A}
 \; \big[ \tilde A(\vec{r}_o)\; {h}(\vec{r}_i,\vec{r}_o)\big]^N\Big|^2
\labell{coherentimaging}\;,
\end{eqnarray}
which, contrary to Eq.~(\ref{nfold}), cannot be obtained through
post-processing of $P_1$, and which generalizes coherent
imaging~(\ref{conventional}) to $N$-photon detection. The point-spread
function that governs spatial resolution is now ${h}^N$---which is
narrower than $h$---so that when $A(\vec{r}_o)\simeq\tilde
A(\vec{r}_o)$ there is an enhancement in resolution over the Rayleigh
bound. More generally, even if $A$ and $\tilde A$ differ
significantly, one can still beat the Rayleigh bound if $N$ is
sufficiently large and $D_i/R\gg {k}/\Delta k_t$, see
Figs.~\ref{f:drago}(d) and (e).

An analogous generalization for incoherent imaging is
obtained by replacing the state Eq.~(\ref{oak}) with an incoherent mixture
of focused Fock states, i.e., $\rho = \int \tfrac{d^2\vec{r}_o}{\cal
  A} |N\rangle_{\vec{r}_o}\langle N|$. In this case
Eq.~(\ref{coherentimaging}) becomes
\begin{eqnarray}
  P_N^{(\text{inc})}(\vec{r}_i)\simeq \xi^N \int
  \tfrac{d^2\vec{r}_o}{\cal A}\;|\tilde A(\vec{r}_o)|^{2N}
  |{h}(\vec{r}_i,\vec{r}_o)|^{2N}
\labell{postsel}\;,
\end{eqnarray}
which generalizes Eq.~(\ref{inco}) to 
$N$-photon detection. 
 The corresponding 
resolution enhancement  is shown in Fig.~\ref{f:logoincoer}.

The states employed in Eqs.~(\ref{oak}) and (\ref{postsel})
are highly sensitive to loss.  Nevertheless, $N$-fold incoherent
imaging can be realized with loss-resistant light sources. Suppose we use an incoherent mixture of coherent states that  randomly illuminate all points on the object:  $\sigma= \int \tfrac{d^2\vec{r}_o}{\cal A} \;|\alpha\rangle_{\vec{r}_o} \langle\alpha|$, where $|\alpha
\rangle_{\vec{r}_o} \equiv \exp[ \alpha b^\dag (\vec{r}_o) - \alpha^*
b(\vec{r}_o) ] |0\rangle$.  Equation~(\ref{postsel})
still applies with an extra multiplicative factor of 
${|\alpha|^{2N}}/{N!}$ to account for the Poissonian photodetection statistics of
coherent states.  The state $\sigma$
can be prepared by shining a highly-focused laser beam 
on the object, one point at the time. This state is highly robust to loss, because loss parameter $\mu$ just takes $|\alpha\rangle$ into
$|\sqrt{\mu}\alpha\rangle$. Hence an arbitrary amount
of loss can be tolerated---without sacrificing resolution---simply by 
increasing $|\alpha|$.  

\begin{figure}[htb]
\begin{center}
{\footnotesize a}  \epsfxsize=.46\hsize\leavevmode\epsffile{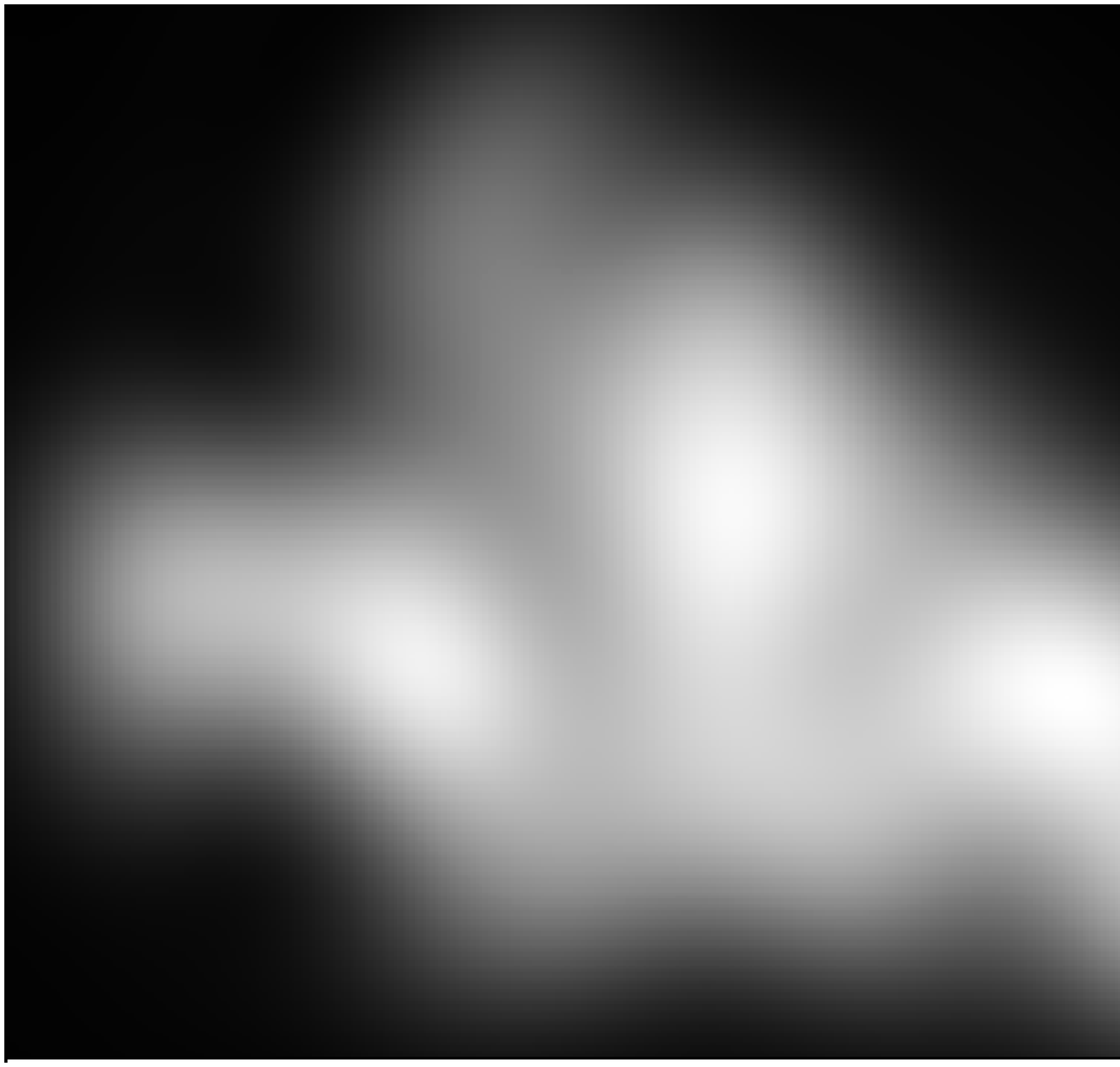}
{\footnotesize b}  \epsfxsize=.46\hsize\leavevmode\epsffile{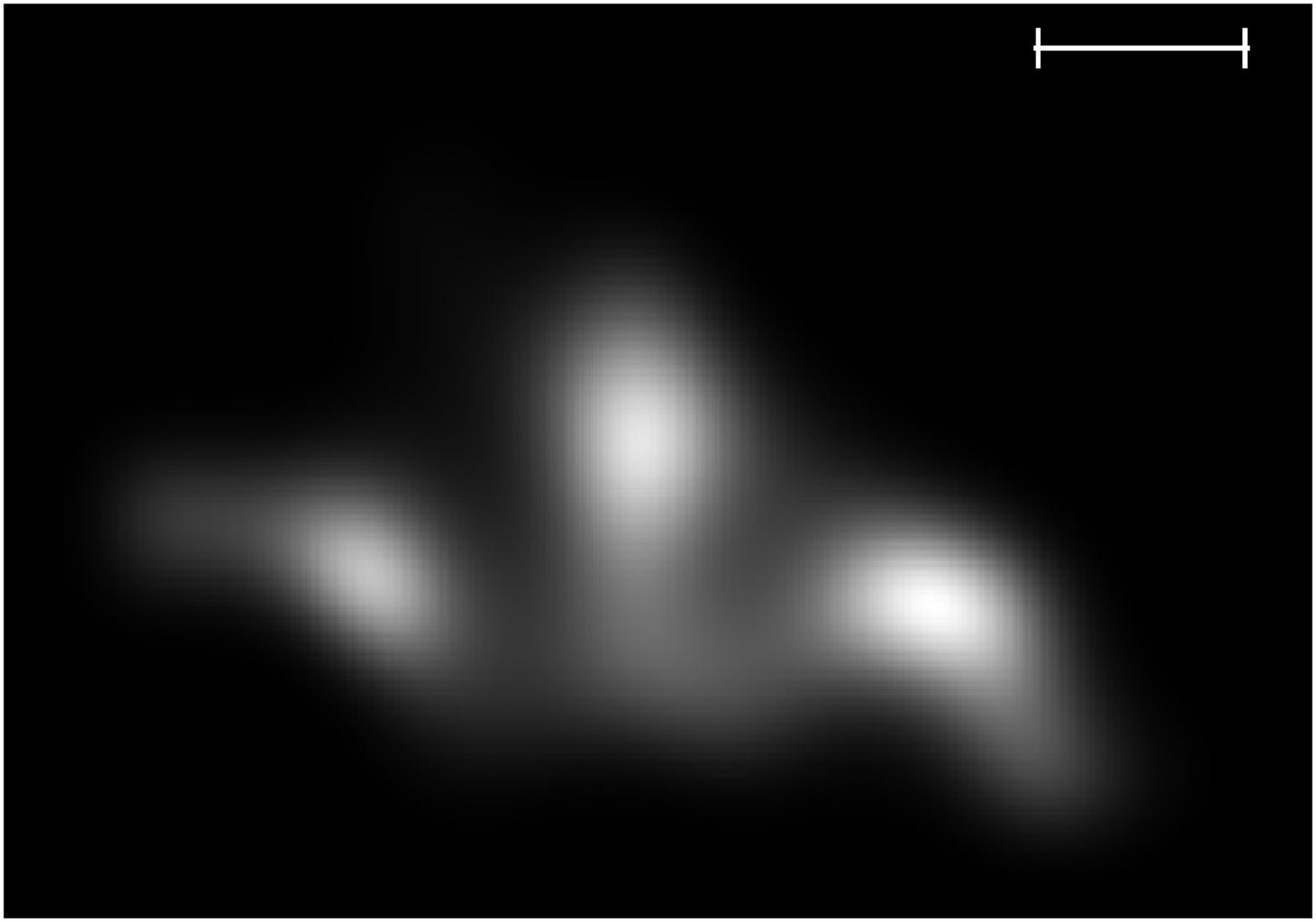}
\end{center}
\begin{center}
{\footnotesize c}  \epsfxsize=.46\hsize\leavevmode\epsffile{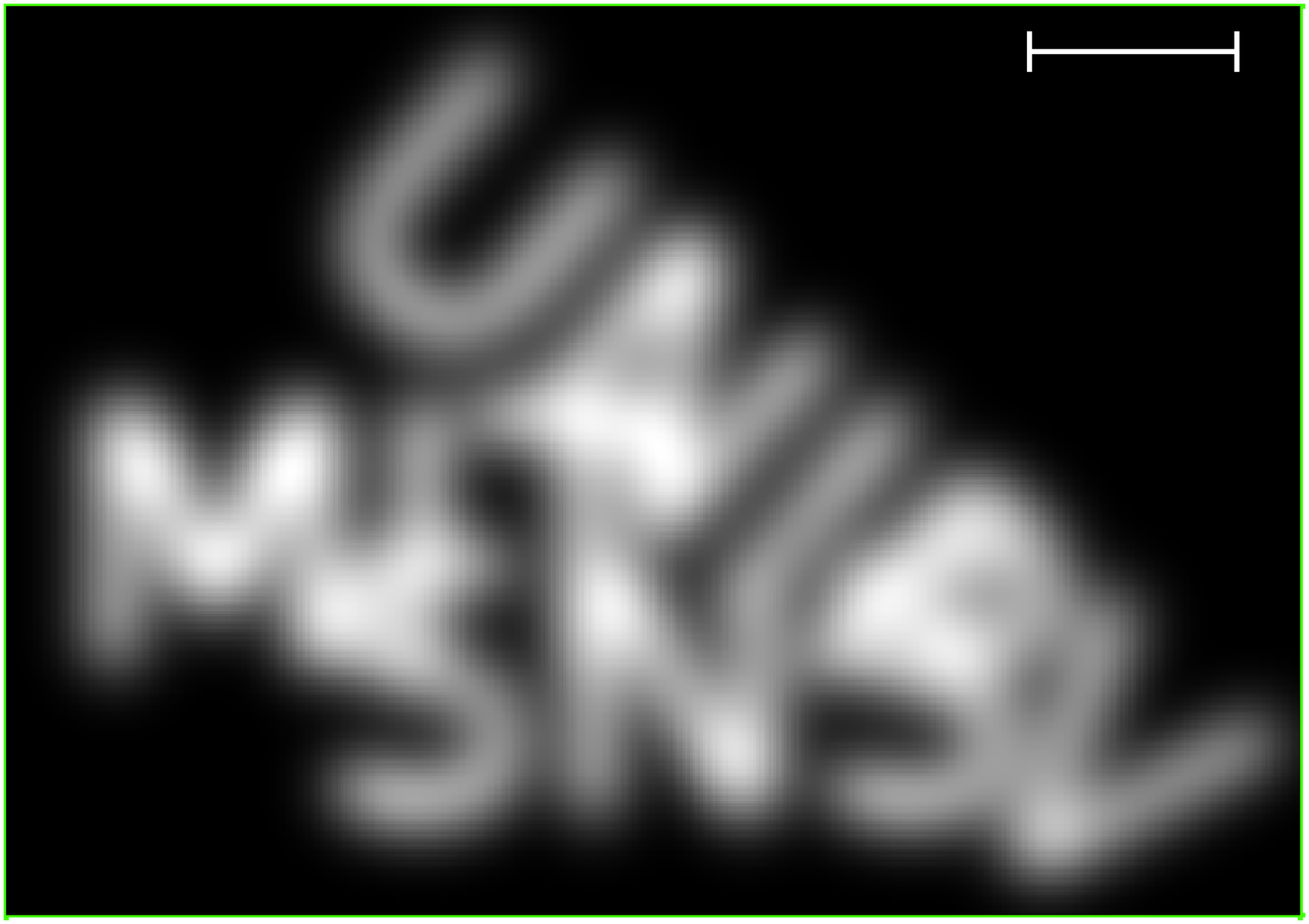}
{\footnotesize d}  \epsfxsize=.46\hsize\leavevmode\epsffile{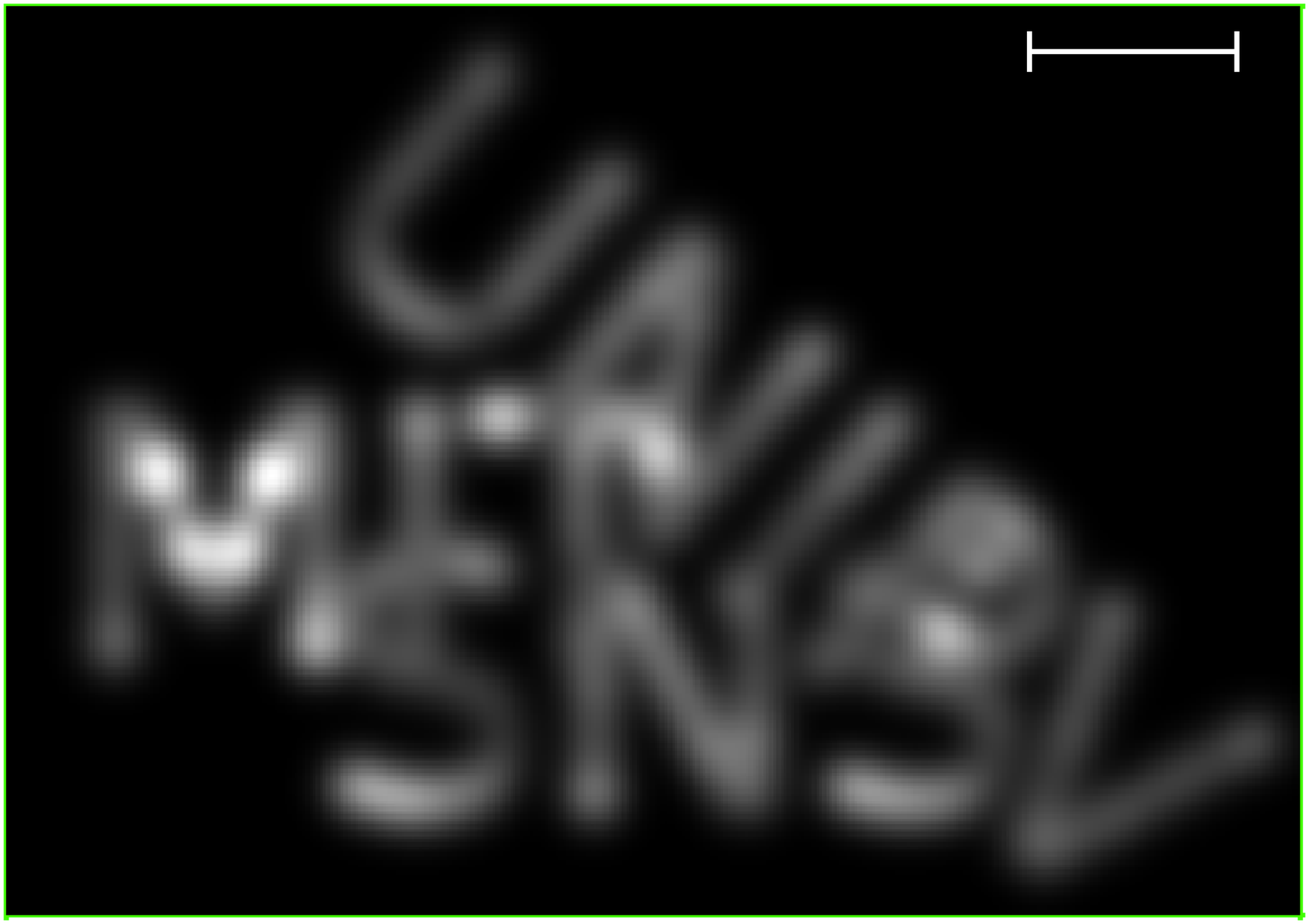}
\end{center}
\caption{Comparison of conventional and standard quantum limit incoherent imaging of the object shown in 
  Fig.~\ref{f:drago}(a) with the same $D_o/R=250$,
  $m=1$, and $k=6000$.  (a)~and (b)~Conventional images from
  Eq.~(\ref{nfold}) with $N = 1$ and 5, respectively.  The images are  featureless blurs, because the
  Rayleigh bound $x_R$ is unable to resolve any object details.
  (c) and (d)~Reconstruction via Eq.~(\ref{postsel}) using the state
  $\sigma$ for $N=5$ and 10, respectively, with $\Delta k_t=600$. An obvious increase in resolution is
  seen.}
\labell{f:logoincoer}
\end{figure}

The improved resolution afforded by the procedures detailed above can be
roughly estimated by gauging by narrowing of the point-spread
function ${h}$ that results from taking its $N$th power.  This can be done,
for instance, by evaluating the radius $x_R(N)$ that contains 84$\%$
of the area under ${\rm somb}^{2N}$ in the plane.  Numerical
analysis shows that $x_R(N)/x_R\sim 1/\sqrt{N}$, which suggests a
standard quantum limit~\cite{review} for imaging.  This should be
taken only as a rough estimate, as $x_R(N)$ is also the radial
dimension of a point-like object imaged using the post-processing
strategy of Eq.~(\ref{nfold}).  For more extended objects, the actual
resolution enhancement will also depend on $\Delta k_t$. The
$1/\sqrt{N}$ scaling exposes the classical nature of this enhancement:
the same effect can be attained by averaging the arrival positions of
$N$ photons at the image plane. This is surely advantageous over
$N$-photon detection in many situations, but it is
impractical for lithography or
film photography, and it seems impossible to
classically reproduce the coherent imaging case of
Eq.~(\ref{coherentimaging}). In addition, from general
principles~\cite{review,metrology} one would expect the ultimate bound
(a Heisenberg limit for imaging) to have $1/N$, scaling,
i.e., a resolution $\sim$$x_R/N$, which is not achievable with classical
strategies.

\paragraph{Heisenberg limit:--}The Heisenberg $1/N$ scaling can be
obtained by treating the $N$ photons as a single entity of $N$-times
higher frequency.  This situation can be simulated, at least in
principle, by inserting immediately in front of the lens a screen
divided into small sections each of area $s_F$ such that if less than
$N$ photons reach one section, they are absorbed, otherwise they are
coherently transmitted.  Such a screen does not currently exist, but in
principle one could be built, e.g., by using doppleron
materials~\cite{doppleron}.  Then, if the object is 
illuminated by the focused coherent states described above, only $N$
photons that originate at $\vec{r}_o$, successfully transit the screen
within one of its area-$s_F$ segments, and get detected at $\vec{r}_i$
can can contribute to the image at that point. In this case, the
operators $[{E}_i^{(+)}(\vec{r}_i,t)]^N$ of the $N$-photon absorption
probability~(\ref{coherentimaging}) are approximately~\cite{NOTENEW1}
\begin{eqnarray}
&&\!\!\!\!\!\![{E}_{i}^{(+)}(\vec{r}_i,t)]^N \simeq \gamma
 \int \tfrac{d^2\vec{r}_o}{\cal A}\:[\tilde A(\vec{r}_o)]^N\;
  h_{N}(\vec{r}_i,\vec{r}_o) \; \Big[b(\vec{r}_o)\Big]^N
\;.\nonumber\\ && \labell{enne}
\end{eqnarray}
Here: $h_N$ is obtained from Eq.~(\ref{acca}) by replacing ${k}$ with
$N{k}$, i.e., $h_N$ is the point-spread function for photons having
$N$-times higher frequency than the illumination; and $\gamma$
accounts for the spatial resolution of the doppleron screen, i.e., it
is of order $(\tfrac{s_F}{ \pi R^2})^N$. Equation~(\ref{enne})
describes the absorption of $N$ frequency-$\omega$ photons that
originated near $\vec r_o$ and then propagated through the imaging
apparatus as if they were a single frequency-$N\omega$ photon.  It
gives rise to coherent and incoherent images that are formally
equivalent to those of Eqs.~(\ref{conventional}) and (\ref{inco}) for
a light beam of wave number $Nk$, thus realizing the Heisenberg limit
of an $N$-fold resolution improvement over the Rayleigh bound, see
Fig.~\ref{f:drago}(f).  This method shows that Heisenberg-limited
resolution can be obtained using classical light, albeit with an even
worse efficiency than the $N$-photon detection methods given above.

This research was supported in part by the W. M. Keck Foundation for Extreme Quantum Information Theory and by the DARPA Quantum Sensors Program.


\begin{references}
\bibitem{nota1}The function $A$  measures the attenuation of the
 impinging light due to the object. Here we assume $A$ to be
  independent of $\vec k$; the generalization to a wave-vector
  dependent $A$ is straightforward.
\bibitem{DEFB}It is given by $b(\vec{{r}_o})\equiv\int \tfrac{d
    \omega}{\sqrt{\Delta \omega}} \int \tfrac{d^2\vec k_{t}}{\sqrt{\pi
      \Delta k_t^2}} \;e^{i\vec k_{t} \cdot\vec{r}_o}a({\vec k})$,
  where the frequency integration is performed over a small bandwidth
  $\Delta \omega$ around the mean frequency $\omega = kc$ and the wave-vector integration  is performed over $|k_{t}|\leqslant\Delta k_t$.  The exact
  expression for ${\cal M}$ is $\int {d^2\vec{r}_o} \int
  {d^2\vec{r}'_o} \; {F_{\Delta k_t^2}^N(|\vec{r}_o -
    \vec{r}_o'|)}/{{\cal A}}^N$, which, for highly localized modes
  $\pi \Delta k_t^2 {\cal A} \ll1$, gives ${\cal M} \simeq \tfrac{16
    \pi {\cal A}}{\Delta k_t^2}$.
\bibitem{NOTENEW1} Use the screen-to-lens 
  transfer function to write $[{E}_i^{(+)}(\vec{r}_i,t)]^N$ as a
  convolution of the products
  ${E}_\ell^{(+)}(\vec{r}_\ell^{(1)},t)\cdots{E}_\ell^{(+)}(\vec{r}_\ell^{(N)},t)$,
  with ${E}_\ell^{(+)}(\vec{r}_\ell,t)$ being the electric field on
  the screen.  The screen removes all contributions save those for
  which the $\vec{r}_\ell^{(j)}$ fall in the same area-$s_F$ region.
  Equation~(\ref{enne}) follows by writing the resulting operator as a
  convolution of ${E}_o^{(+)}(\vec{r}_o^{(1)},t)\cdots
  {E}_o^{(+)}(\vec{r}_o^{(N)},t)$ through the object-to-lens
  transfer function, and by noticing that for focused sources it can
  be approximated by $[b(\vec{r}_o^{(1)})]^N$.
\end{references}
\end{document}